%% file: murat_multi-rat_false_base_station_detector.tex
\documentclass[conference]{IEEEtran}
\usepackage{graphicx}
\usepackage{listings}
\usepackage{textcomp}
\usepackage{url}
\usepackage{cite}
\usepackage [english]{babel}
\usepackage [autostyle, english = american]{csquotes}
\MakeOuterQuote{"}
\begin{document}

\input{parts/0.title}
\input{parts/0.abstract}
\input{parts/1.introduction}
\input{parts/2.background}
\input{parts/3.murat_description}
\input{parts/4.lab_experiments}
\input{parts/5.operator_trial}
\input{parts/6.standardization}

\input{parts/7.effectiveness_limitations}
\input{parts/8.future_work}
\input{parts/9.conclusion}
\input{parts/91.acknowledgement}

\bibliographystyle{IEEEtran}
\bibliography{parts/91.references}
\end{document}

%% file: parts/0.title.tex
\title{Murat: Multi-RAT False Base Station Detector}

\author{
    \IEEEauthorblockN{Prajwol Kumar Nakarmi}
    \IEEEauthorblockA{Ericsson Research Security\\
    prajwol.kumar.nakarmi@ericsson.com}
    \and
    \IEEEauthorblockN{Mehmet Akif Ersoy}
    \IEEEauthorblockA{Ericsson Research Security\\
    mehmet.akif.ersoy@ericsson.com}
    \and
    \IEEEauthorblockN{Elif Ustundag Soykan}
    \IEEEauthorblockA{Ericsson Research Security\\
    elif.ustundag.soykan@ericsson.com}
    \and
    \IEEEauthorblockN{Karl Norrman}
    \IEEEauthorblockA{Ericsson Research Security\\
    karl.norrman@ericsson.com}
}

\author{
    \IEEEauthorblockN{
        Prajwol Kumar Nakarmi\IEEEauthorrefmark{1},
        Mehmet Akif Ersoy\IEEEauthorrefmark{2}, 
        Elif Ustundag Soykan\IEEEauthorrefmark{2} and
        Karl Norrman\IEEEauthorrefmark{1}}
            
    \IEEEauthorblockA{
        Ericsson Research Security\\
        \IEEEauthorrefmark{1}Stockholm, Sweden\\
        \IEEEauthorrefmark{2}Istanbul, Turkey\\
        \{prajwol.kumar.nakarmi, mehmet.akif.ersoy, elif.ustundag.soykan, karl.norrman\}@ericsson.com}
}


\maketitle

%% file: parts/0.abstract.tex
\begin{abstract}
In recent years, there has been an increasing interest in false base station detection systems. Most of these rely on software that users download into their mobile phones. The software either performs an analysis of radio environment measurements taken by the mobile phone or reports these measurements to a server on the Internet, which then analyzes the aggregated measurements collected from many mobile phones. These systems suffer from two main drawbacks. First, they require modification to the mobile phones in the form of software and an active decision to participate from users. This severely limits the number of obtained measurements. Second, they do not make use of the information the mobile network has regarding network topology and configuration. This results in less reliable predictions than could be made. We present a network-based system for detecting false base stations that operate on any 3GPP radio access technology, without requiring modifications to mobile phones, and that allows taking full advantage of network topology and configuration information available to an operator. The analysis is performed by the mobile network based on measurement reports delivered by mobile phones as part of normal operations to maintain the wireless link. We implemented and validated the system in a lab experiment and a real operator trial. Our approach was adopted by the 3GPP standardization organization.
\end{abstract}

%% file: parts/1.introduction.tex
\section{Introduction}
\label{intro}
Mobile networks provide a fundamental part of our every-day communication that deliver wireless internet access and services on a global scale. They are also increasingly considered as critical infrastructure \cite{criticalUK, criticalUSA}. Therefore, it is crucial to protect them from potential attacks. An important attack vector is the radio interface between the mobile phone and the base stations of a mobile network. Radio devices making use of this attack vector by impersonating a mobile network's base station towards a mobile phone is often referred to as \textit{false base stations}. These devices come in many flavors, some of which also impersonate a mobile phone towards the mobile network. 

Attacks that can be performed using false base stations broadly fall under categories as follow - privacy compromise of mobile phone usage, denial of service (DoS) on mobile phones, DoS on mobile network, and frauds.  The efficacy of these attacks, however, greatly varies between different generations of mobile networks as each newer generation becomes more resilient than earlier. Nevertheless, due to many reasons like legacy networks that have over lived their time and interworking between wide variety of networks, the mobile network industry is not fully protected against all types of attacks from false base stations. Detection and protection against false base stations is therefore an important topic for mobile network industry and society as a whole. 

Over the past couple of years, a number of systems for detecting false base stations have been proposed and prototyped. Most of these implement a data collection capability in the mobile phone, which either perform analysis on the collected data in the mobile phone or send the collected data to a central server for analysis. We refer to the first type as User Equipment (UE)-based, e.g., \cite{AIMSICD, Snoopsnitch, cryptophone}, and the second type as crowd-sourced detectors, e.g.,  \cite{Ney2017SeaGlassEC, fade, Li2017FBSRadarUF}. What is common for these types is that they determine whether a false base station is present by making use of the view of the network from a mobile phone's perspective. This view is by its very nature limited in comparison to the view from a network's perspective. Not only does a mobile network have knowledge of the global state of the system, whereas a mobile phone only has knowledge of its local state, but the mobile network also has more knowledge of a mobile phone's state than the mobile phone itself. The mobile phone only has sufficient knowledge of its state to operate when being commanded by the network to take certain actions. 

Another drawback of these detection systems is that they require modified mobile phones. This means that end users must download and run a special application to collect and analyze measurements or report the measurements to a central server on the internet for analysis. This severely reduces the number of measurements accessible for analysis. 

A few false base station detectors are network-based that rely on information collected by the mobile network. However, they do either require a pre-existing network monitoring infrastructure \cite{Dabrowski2016TheMS} or focus on a single 3GPP Radio Access Technologies (RAT) \cite{Steig2016ANB}.

In this paper, a system addressing the issues above -- \emph{Murat} -- is proposed. Murat is a network-based system for detecting false base stations that can operate on multiple 3GPP RATs, without requiring any modification to mobile phones or separate monitoring infrastructure. 

Murat builds upon the blog post \cite{earlierblog} that is our earlier work. This paper extends the blog post in three ways. Firstly, it enriches the content of the blog post with background and discusses related works. Secondly, it elaborates the system description which is done only briefly in the blog post. Lastly, this paper adds how the system was validated in a real operator trial.

 Murat is designed as a network functionality making use of the global state incorporating information about all connected mobile phones' states, as well as knowledge about the mobile network state and its deployment, configuration and execution history. This gives an information advantage over previously proposed systems. 

The core idea underlying our system is to make use of this information advantage by comparing the view of mobile phones connected to a mobile network with the views that the mobile network intends the mobile phones to have. If there is a discrepancy between certain parts of the mobile phones' views and what the network expects, it is an indication that a false base station may be present. This strategy is made possible by the fact that mobile phones regularly report their local views to the mobile network as part of normal operation. Without this reporting, the network would not be able to maintain the wireless connections to the mobile phones. 

To give a simple but illustrative example of the core idea, assume the mobile network uses five legitimate base stations configured to serve an area. Each of these base stations needs to have unique identifiers, e.g., to facilitate handovers of mobile phones as they move in and out of radio coverage of different base stations.  As the mobile phones move in the area, they measure signal strengths for different base stations and report these measurements to the network. While it is difficult for a single mobile phone, or even a collaborating set of mobile phones, to determine whether any of the measured base stations are legitimate or false, the mobile network has information about which base stations are configured to operate in the area, what combinations of signal strengths can be expected to be measured by a mobile phone, which identifiers should exist in the area and so on. If these parameters deviate too much and too frequently with respect to some thresholds, it indicates that a false base station may be present. Depending on parameter and threshold choices, false positives and false negatives may occur, so system tuning is required for optimal performance.

We evaluated the effectiveness and complexity of deploying Murat by performing a lab experiment and a real operator trial. The lab experiment included a controlled setup with multiple base stations using different 3GPP RATs (2G and 4G) and a mobile phone in a Faraday's cage. This showed the important aspect that even though a mobile phone is connected to a specific 3GPP radio access technology, say 4G, it still can be configured to report measurements for other 3GPP radio access technologies, say 2G. This makes it possible to deploy Murat covering one radio access technology and still be able to detect false base stations operating on other radio access technologies.  The operator trial included running Murat in a real operator's 4G network with a planted false base station built using publicly available tools. Both the lab experiment and operator trial showed efficacy of Murat on seamlessly integrating to mobile networks and being able to flag suspicious discrepancies. We proposed our approach to the 3GPP standardization organization, which develops standards for mobile networks. It has been adopted as part of the 5G specifications. 

We summarize our main contributions as follows
\begin {itemize}
\item We present a network-based system for detecting false base stations that operate on any 3GPP radio access technology (RAT), without requiring any modification to mobile phones. 

\item We verified detection properties of the system both in a controlled lab experiment and in a real operator trial. We also share our insights from them in this paper that provide guidance for real world deployments. 
\item We proposed the approach to 3GPP, and they have adopted it in the mobile network standards.
\end {itemize}

The paper is structured as follows. We provide some background information about mobile networks, false base stations, existing countermeasures, and what has been done in the detection area in the Section \ref{bg}. We introduce our system, multi-RAT false base station detector called Murat in Section \ref{murat}. We describe the lab experiment in Section \ref{labex} and the operator trial in Section \ref{optr}. We continue with describing impact of our work on standardization in Section \ref{std}. Effectiveness and limitations of Murat are discussed in Section \ref{eff}. We provide pointers to future research in this topic in Section \ref{fr}, and we conclude the paper in Section \ref{conc}.    
 

%% file: parts/2.background.tex
\section{Background}
\label{bg}
\subsection{Mobile Networks}
\label{bg_mn}
\begin{figure}[h]
\centering
\includegraphics[width=3.5in]{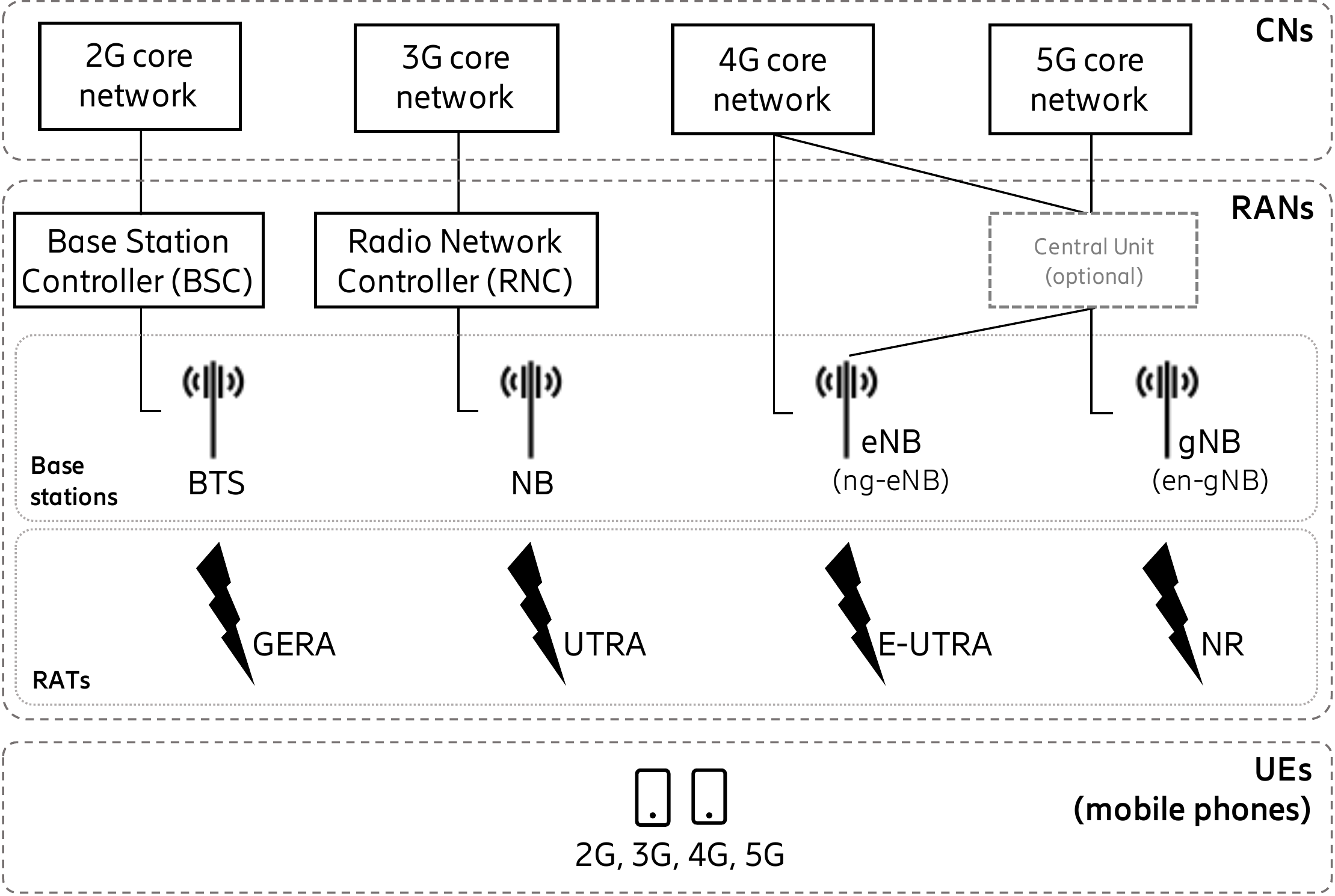}
\caption {Simplified overview of mobile networks} 
\label{fig:networks} 
\end{figure}

Mobile networks have evolved from 2G in 1991 to 3G in 2001 to 4G in 2008. The most recent generation is the 5G in 2018, with the first commercial deployments of 5G happening in 2020. A simplified overview of different generations of mobile networks is shown in Fig. \ref{fig:networks}. On a high level, a mobile network consists of UEs which refer to mobile phones that we use, Radio Access Network (RAN) which are network entities providing wireless radio access to the UEs, and Core Network (CN) which are the set of network functions that among other things do subscription handling, session and mobility management, and packet routing. 

3GPP \cite{3gpp} is an organization that standardizes mobile networks, including their architecture, protocols, and security. Then, mobile network operators plan and deploy the networks according to their needs, e.g., how many and what combination of base stations to install in an area. The actual deployment options of each generation, especially when it comes to 5G, and interworking between each other is complex and we will not go into their details, rather we briefly mention some RAN aspects that are related to this paper.

Historically, different generations of RAN offer radio access via a radio access technology (RAT) specific to that generation, i.e., 

\begin{itemize}
\item 2G RAT: GSM EDGE Radio Access (GERA)
\item 3G RAT: Universal Terrestrial Radio Access (UTRA)
\item 4G RAT: Evolved-UTRA (E-UTRA)
\end{itemize}
In 5G, however, two types of RAT can co-exist, one 4G RAT (E-UTRA) and another a new one, i.e.,
\begin{itemize} \item 5G RAT: New Radio (NR). \end{itemize}

The RAN uses base stations to offer these RATs. They are known as Base Transceiver Station (BTS) in 2G, NodeB (NB) in 3G, Evolved NB (eNB) in 4G, and next-Generation NB (gNB) in 5G. There also exist ng-eNB (Next-Generation eNB) that connects to a 5G core network and en-gNB (E-UTRA New Radio gNB) that connects to a 4G core network. These base stations support one or more cells, a so-called cell being the smallest coverage area in which the base stations serve the UEs.

\subsection{False Base Station Attacks}
\label{bg_fbs}
\begin{figure}[h]
\centering
\includegraphics[width=3.5in]{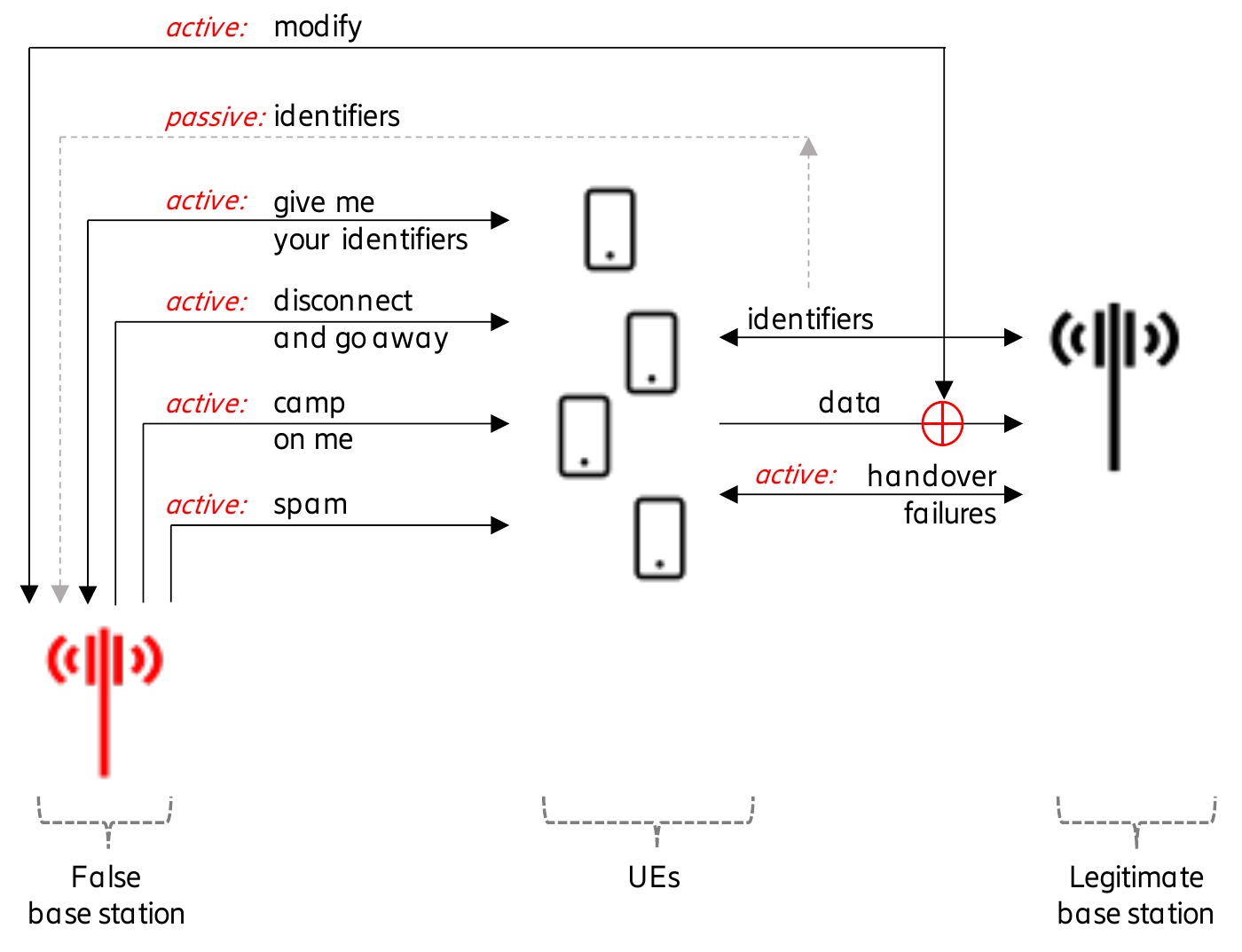}
\caption {Logical illustration of false base station attacks} 
\label{fig:fbsd} 
\end{figure}

False base station is a broad name for a radio device which sets out to impersonate a legitimate base station. Although the name says, "base station", its attack capabilities have outgrown to also impersonate UEs towards the mobile network. It is also known by other names such as IMSI catcher, Stingray, rogue base station, or cell site simulator. A logical illustration of false base station attacks is shown in Fig. \ref{fig:fbsd}.

One of the main attacks relates to privacy of users, in which an attacker either passively eavesdrops users' identifiers from the radio interface or actively obtains them by communicating with the UEs. The attacker then uses those identifiers to identify or track the users \cite{Shaik2016PracticalAA, Borgaonkar2018NewPT, Hussain20195GReasonerAP, Hussain2019PrivacyAT}. The attacker might also try to fingerprint user traffic \cite{Kohls2019LostTE}. Under stringent assumptions, a resourceful attacker may also exploit implementation flaws \cite{Rupprecht2020CallMM} or vulnerabilities in application layer protocols like Domain Name System (DNS) and Internet Control Message Protocol (ICMP) by altering carefully chosen parts of the data in the radio interface \cite{rupprecht-19-layer-two, Rupprecht2020IMP4GTIA}. 

Another set of attacks relates to denial of service (DoS) on UEs and/or mobile network. The attacker may do this by using certain messages that the UEs and the network accept without authentication \cite{Shaik2016PracticalAA, Kim2019TouchingTU}. The attacker may also create favorable radio conditions so that the UEs keep camping on the false base station, thereby, being cut off from all incoming communications from legitimate base stations. Radio conditions created by the attacker may also trigger certain events in the legitimate network like handover failures. Because of these events, some implementations in the network may take disruptive steps such as barring even the legitimate base stations, thereby, triggering service disruption \cite{Shaik2018OnTI}. 

Fraud attacks with financial motive may send spam or advertising SMS messages to UEs \cite{Li2017FBSRadarUF} or even try to impersonate them \cite{Chlosta2019LTESD, Rupprecht2020IMP4GTIA}. There could also be non-financial motive in which the attacker may poison UE's location \cite{Hussain2018LTEInspectorAS} or send public warning messages to create panic in public \cite{Lee2019ThisIY, Yang2019HidingIP}.

While false base stations could be used for good of the society, such as tracking down criminals or locating lost children, they could also be used to harm the functioning of the society with dire consequences like unauthorized surveillance, communication sabotage, unsolicited advertising, or even physical harms. Further, with the increase of the connectivity over varieties  of devices, false base stations may not just target the mobile phones. Knight et. al \cite{hacking} mention using false base stations in their book related to hacking connected cars. 

It is also getting cheaper and easier to setup a false base station. On the hardware side, the overall price required varies around couple of thousands USD \cite{Dabrowski2014, Kohls2019LostTE, Borgaonkar2018NewPT}. On the software side, open source stacks are getting mature and popular \cite{srsLTE, oai}. 

Therefore, increased incentive for attacker due to ever-growing connectivity and increased feasibility of deploying a false base station are main reasons that false base station attacks are more important now than ever. It has rightly got more attention in all fronts, media, hackers conferences, academia, standardization bodies, governments, law enforcement agencies, vendors, and operators.

\subsection{Existing Countermeasures}
\label{bg_ec}
While 2G remains most vulnerable among all generations, several attacks have become impractical or the level of difficulty for attackers has significantly increased with newer generation of mobile networks. Especially, 5G has been designed with significant enhancements in terms of privacy and security against false base stations. For example: encryption of permanent identifier (known as SUPI in 5G, equivalent to IMSI in earlier generations) makes it more difficult for false base stations to track users by eavesdropping this identifier over the radio interface; and integrity protection of user traffic enables detection of data alteration by false base stations in the radio interface  \cite{3gpp33501, nakarmiblog}. Nevertheless, attacks from a false base station attack are not completely eradicated yet. Some attacks are possible simply because newer networks are required to interwork with 2G networks.  Also, attacks may be possible merely because some security features were not activated. 3GPP is currently assessing if and what type of further enhancements can be done in 5G in terms of new protection and detection features \cite{3gpp33809}.

The feature that we will deal in detail through rest of the paper is detection of false base stations in order to make it significantly harder for false base stations to remain stealthy. Even though protection features are in place, detection is still important. It is so because detection is not alternative to protection, rather is complementary, and forms a fundamental function of cybersecurity \cite{NIST}.

\subsection{Related Work on False Base Station Detection}
\label{bg_rw}
We briefly mentioned existing false base station detection systems and their drawbacks in the introduction; here we give more details about them and their limitations. We classify them according to the methods they used. First ones are UE-based detectors that do the detection directly in the UEs (i.e., mobile phones). Second are crowd-sourced detectors that collect information from large number of UEs or dedicated sensors and do detection in some central server independent of mobile network operators. Third ones are network-based detectors which do the detection directly in the mobile network controlled by the mobile network operators.

\subsubsection{UE based Detectors}
\label{bg_ue}
Brenninkmeijer \cite{brenninkmeijer_2016} and Borgaonkar et al. \cite{Park2017WhiteStingrayEI} evaluated detector applications by setting up a test network and comparing how those applications behave. In \cite{brenninkmeijer_2016}, AIMSICD \cite{AIMSICD}, SnoopSnitch \cite{Snoopsnitch} and Cell Spy Catcher \cite{SKIBAPPS} are analyzed. In addition to these, \cite{Park2017WhiteStingrayEI} also evaluated Darshak \cite{darshak} and GSM Spy Finder \cite{GALAN}. Both works use Android based UEs, software defined radio application and USRP hardware for testing environment. Brenninkmeijer found that while those detectors provide some level of alarm to inform about local network configuration changes, they do not perform well even when they have high operational privileges on the UE. Therefore, the evaluated detectors were not recommended for public. Authors in \cite{Park2017WhiteStingrayEI} states that the lack of root access prevents the detectors from detecting some of the attacks, and in general none of the detectors were perfect in terms of identifying false base stations. In \cite{Abodunrin2015DetectionAM}, Dare also proposed an approach that relies on SnoopSnitch and AIMSICD. Measurements were performed first on 3G mode and then in 2G mode by walking around an area while running active tests to identify abnormalities such as unusual power and cell identifiers. The authors say that the detections could not be said to be solid enough and suggest that collaboration is done with mobile network operators. 

Simula Research Laboratory conducted an investigation \cite{SimulaResearch} about the reported use of false base stations in Oslo by a daily newspaper \cite{norway1}. The evaluation was done on the data/alarms obtained  from a specialized UE called Cryptophone \cite{cryptophone} and a measurement hardware/software called Delma. The investigation suggested that the data/alarm obtained from them have shortcomings and therefore were responsible for false alarms.

Alrashede et al. \cite{Alrashede2019IMSICD} proposed using cell fingerprinting based on cell identifiers and locations to detect presence of false base stations. The detection could be done in mobile phones by using some special software and publicly available cell information data. Their proposal was only theoretical.

\subsubsection{Crowd-sourced Detectors }
\label{bg_cd}
Dabrowski et al. \cite{Dabrowski2014} developed a dedicated stationary device and placed it in the field supplied with good antenna. Their device scanned the area in passive mode for cell related fingerprints to detect anomaly. They also developed a UE application based on Android public API facilitating baseband information and built-in GPS receiver with no root privileges. Their results showed that refinement is needed in such cases where GPS reception is bad like underground even though network coverage is good. SITCH \cite{AshWilson} and SeaGlass \cite{Ney2017SeaGlassEC} follow a similar approach offering the sensor devices. Their initiative further became a project called FADe (Fake Antenna Detection Project) \cite{fade} to detect false base stations in Latin America. They used sensors installed into volunteers’ vehicles. The data gathered were aggregated into a city-wide view and analyzed to find anomalies. 

Van Do et al. \cite{Do2015DetectingIU} proposed a solution that uses machine learning to detect abnormal behaviors caused by false base stations on public data set. This study is extended incorporating other machine learning methodologies in \cite{Thuan2016StrengtheningMN} using a signature-based approach with different anomaly indicators like location update, handover use cases and relationship between subscription and UE identifiers. Their experiments are based on publicly available data set from Aftenposten \cite{norway2} aiming to show that there is a potential of applying machine learning techniques.

FBS-Radar \cite{Li2017FBSRadarUF} uses crowdsourced data to detect and geolocate the false base stations. The main threat factor FBS-Radar is dealing with is spam and fraud SMS messages launched by active attackers. The data collection is done by UE and includes suspicious SMS messages and associated meta data, received signal strengths, cell identifier and MAC addresses for WiFi connected UEs. These reports are sent to a cloud for analysis. 

\subsubsection{Network-based Detectors}
\label{bg_nd}
Dabrowski et al. \cite{Dabrowski2016TheMS} discussed techniques from a mobile network point of view, e.g., detection of cipher downgrades, transmission delay, and unusual location identifiers. They used data from a network monitoring infrastructure at a real operator network.   

Steig et al. \cite{Steig2016ANB} utilized measurement reports sent by UEs to 2G network. Their method analyzes the Absolute Radio Frequency Channel Number (ARFCN) and Base Station Identity Code of a neighbor cell (BSIC) from the measurement reports to identify cells belonging to a false base station.

\subsubsection{Limitations of earlier works}
\label{bg_lim}
UE-based detectors are prone to give false positives because a UE or combination of them cannot know the true state of the view of the network at any given instance. A simple example is when the mobile network operator installs a new base station, all UE-based detectors will determine that it is a false base station because it was never seen before. This is an effectiveness issue.

Another drawback of UE-based detectors is that they almost always require modified UEs or privileged root access, which is not common and could be impossible for some users. This means that users must download and run a special application to collect and analyze measurements. Further, operating systems and baseband chips vary a lot among UEs, meaning that the same detector that works in some UEs may not work in others. These severely reduces the number of measurements accessible for analysis. This is a scalability issue.

In case of crowd-sourced detectors, the UEs report the measurements to a central server, e.g., on the internet for analysis. Otherwise, they share the same principle as UE-based detectors and hence suffer from the same effectiveness and scalability issues mentioned above.

Network-based detectors can be expected to perform better when it comes to analysis because whereas UEs only have knowledge of their local state, a mobile network has knowledge of the global state of the system. A limitation, however, with \cite{Dabrowski2016TheMS} is that their system requires an additional pre-existing network monitoring infrastructure that can collect data from various protocols or network points and supply processed data. Such network monitoring infrastructure may not be present or have widely different capabilities among mobile operators, which greatly affects detection mechanisms in \cite{Dabrowski2016TheMS}. Limitation with \cite{Steig2016ANB} is that they only cover 2G RAT and, besides, their result has not been validated in a real operator network.

%% file: parts/3.murat_description.tex
\section{Multi-RAT False Base Station Detector}
\label{murat}
\subsection{Overview}
\begin{figure}[h]
\centering
\includegraphics[width=3.5in]{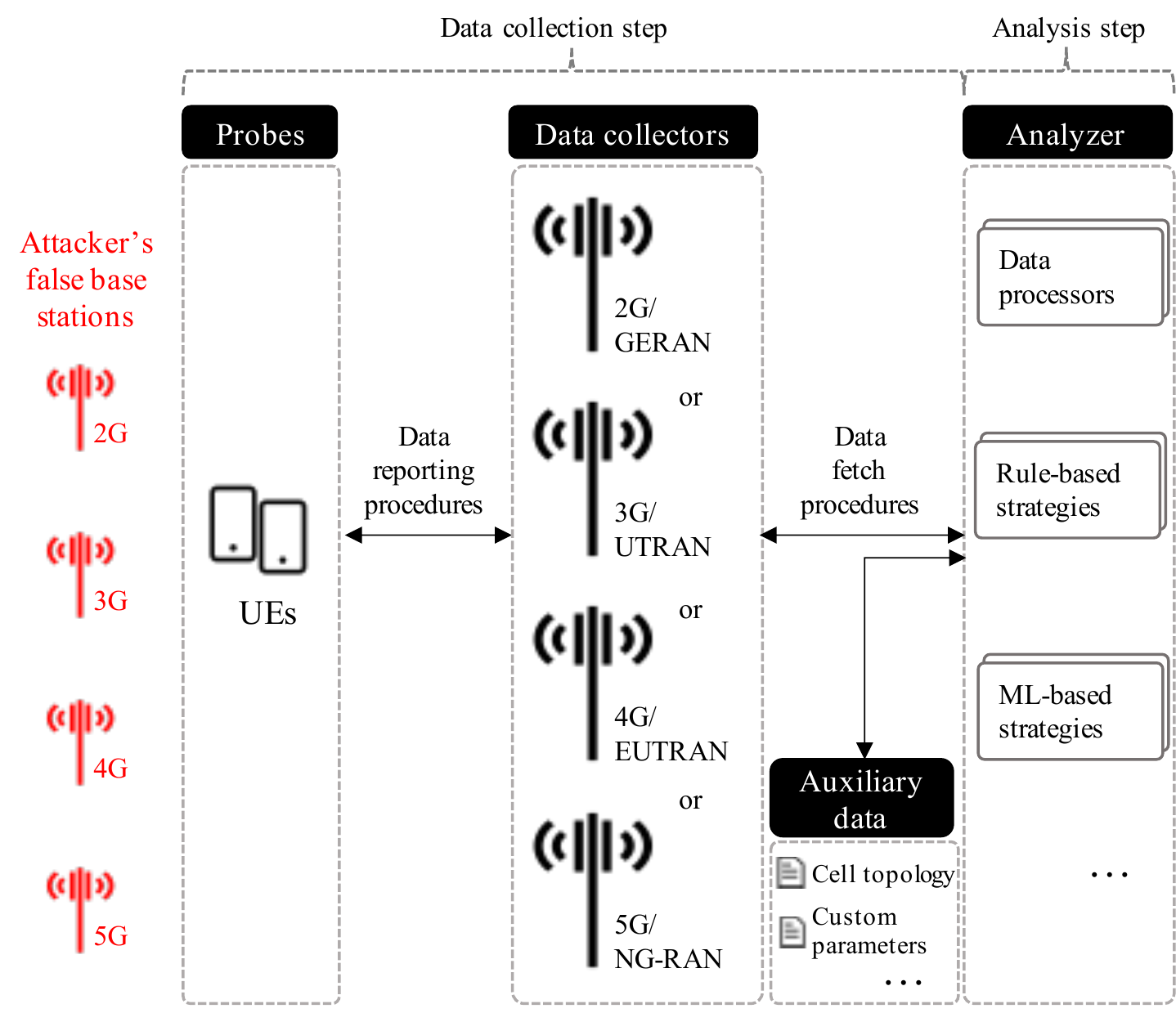}
\caption {Murat, a multi-RAT false base station detector } 
\label{fig:murat} 
\end{figure}

Our multi-RAT false base station detector, which we call Murat, is illustrated in Fig. \ref{fig:murat}. It falls under the category of network-based detector and it builds on what has been briefly described in the blog post by Nakarmi et al. \cite{earlierblog}.

An attacker may operate false base stations for one or more generations of mobile networks and appear as cells with corresponding RATs. In order to detect those false base stations, Murat comprises of four main components as below:

\subsubsection{Probes}
These are ordinary UEs. Murat capitalizes on the fact that a false base station attack is useless to an attacker if there are no UEs around it, and if UEs are present near it, then some of those UEs could be used as probes for data collection without installing any special software on UEs. Depending upon several factors, such as false base station's transmission power, distance from UEs, signal strength of legitimate base stations, and radio conditions, some UEs may fall victim, while others will not. Those UEs, which are sufficiently near the false base station to receive its signals but have not yet fallen victim to it, observe false cells and report them along with other legitimate cells to the operator's legitimate cells. Another fact that Murat exploits is that even ordinary UEs generally support multiple RATs. For example, majority of UEs that work on today's 4G network can also connect to 3G and 2G. The same will likely continue to happen in future, i.e., newer UEs will support multiple RATs.

\subsubsection{Data Collectors}
These are entities in an ordinary RAN which can obtain the main data used in Murat, i.e., measurement report data sent by UEs.  Generally, they are the legitimate base stations belonging to one or more generations which directly engage with UEs and receive reports from them using standard 3GPP procedures. There could be multiple Data collectors to collect measurement reports from different generation/RATs. However, to detect false base stations from a particular generation/RAT, it is not required that legitimate base stations from the same generation/RAT are used as Data collectors. It is so because the actual Probes are UEs that support multiple generations/RATs. Further, some other network functions could also be used as Data collectors even though they do not directly engage with UEs. For example, an operations and maintenance (O\&M) server that manages the RAN could assemble data received by one or more base stations.

\subsubsection{Auxiliary Data}
These are types of data which are other than the measurement reports and can be used in detection of false base stations. They enrich or augment the measurement reports. One example of Auxiliary data is a cell topology data that contains information about base stations present in the operator's mobile network like their location, cells, and RAT types. The cell topology enables the mobile network to compare the UE's view of the mobile network with its own expected view. Another example is files containing customization parameters to be taken into considerations, such as priority timings for detection and thresholds related to signal properties. These parameters enable Murat to adapt or adjust its components when required.  Likewise, there could be other Auxiliary data that enables the mobile network to identify that certain phenomenon is abnormal, such as network event logs containing information of UEs being unreachable, abnormal load in base stations, and problems during cell configurations.

\subsubsection{Analyzer}
This component acts as a brain in Murat to detect false base stations and comprises of multiple functions. One of its functions processes the main data (measurement reports) obtained from Data collectors and the Auxiliary data. The Analyzer is able to take input from several Data collectors and different types of Auxiliary data. This means that even though multiple Data collectors may be used for different generation/RAT, a single Analyzer suffices. Other functions in the Analyzer can apply various strategies, e.g., based on rules or machine learning, which use the processed data and identify if any information contained in it indicate presence of false base stations. New strategies can be added to work on the processed data, e.g., sharing or utilizing threat intelligence to and from another operator's mobile network. It is pertinent to note that the Analyzer can be deployed either as a part of RAN or CN according to mobile network operators' choice, e.g., some operators may choose to deploy the Analyzer directly in base stations while others may choose to use a centralized server in their network. 

In the sections below, we describe how these components of Murat work in two main steps called data collection and analysis.

\subsection{Data Collection Step}
The data collection step in Murat is a combination of procedures between its components that make the required data available to the Analyzer, as below.

\subsubsection{Data Reporting Procedures}
\label{section:meas_report}
\begin{figure}[h]
    \centering
    \includegraphics[width=3.5in]{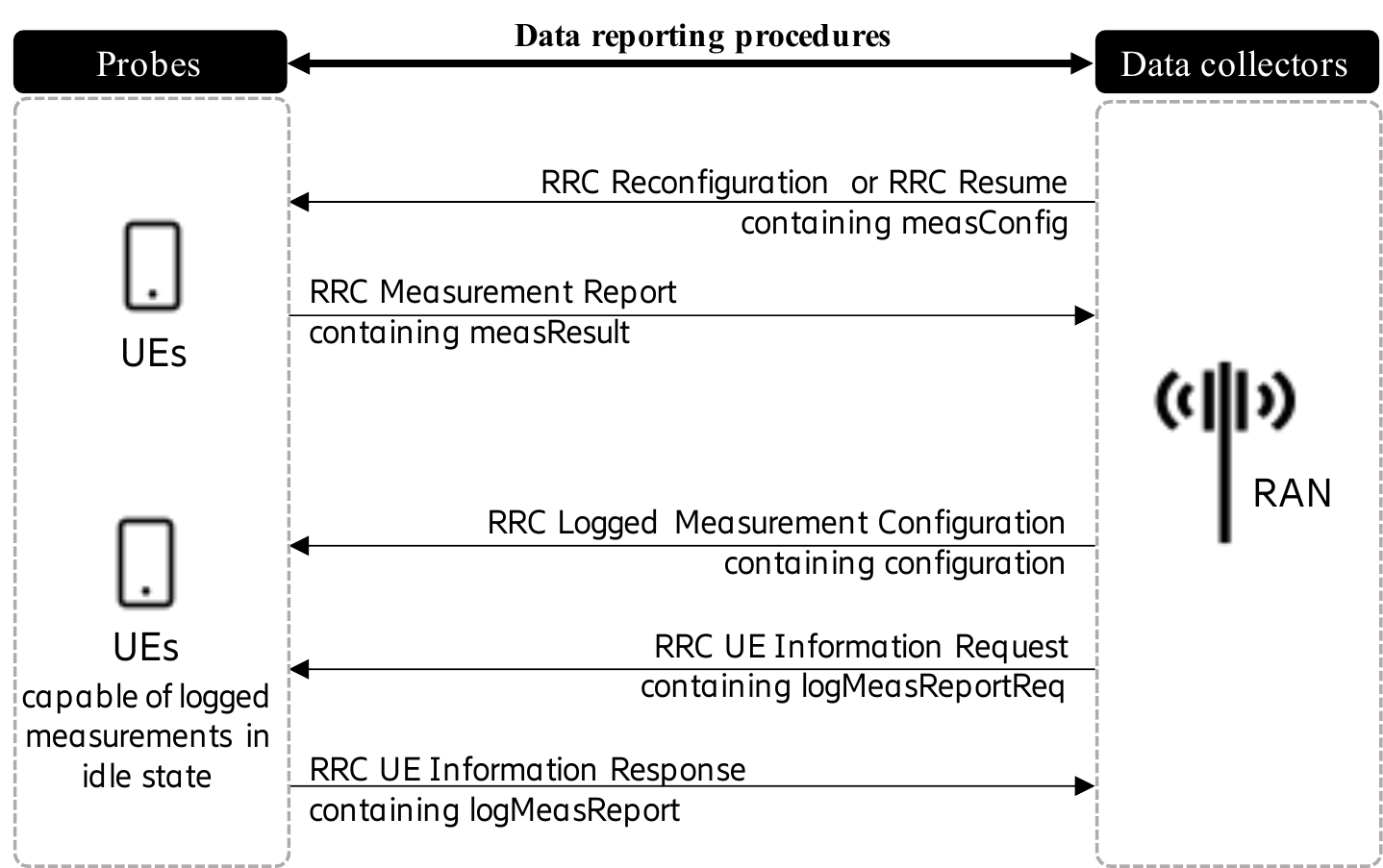}
    \caption {RRC procedures carrying measurement configuration and reports } 
    \label{fig:rrc_proc} 
\end{figure}

Probes (UEs) and Data collectors (RAN) engage in these procedures which take place in the radio interface and belong to standard 3GPP Radio Resource Control (RRC) procedures enabling measurement reporting mechanism. The measurement reporting mechanism is fundamental to all generation of mobile networks and is necessary even for normal operation, for example, it enables the mobile network to decide when radio conditions are such that a UE needs to be served by a different base station, possible even by a different RAT. For brevity, we will stick to describing the measurement reporting mechanism using terminology of 4G and 5G, illustrated in Fig. \ref{fig:rrc_proc}. 

First, the RAN configures UEs in connected state for measurements using RRC messages, so-called reconfiguration and resume. These messages contain measurement configuration that includes many parameters among which we briefly discuss two which are more relevant, namely, measurement objects and reporting configurations. Measurement objects (such as carrier frequency and cell identifiers) are the radio resources on which UE is asked to perform measurements. A 4G network can configure UEs to do measurements on 4G/intra-frequency, 4G/inter-frequency, inter-RAN 5G/NR, 3G, and 2G. A 5G network can configure mobile phones to do measurements on 5G or inter-RAT 4G and 3G frequencies. Measurement objects in 4G cover more RATs than 5G because mobility from 5G is restricted to 4G and 3G. Reporting configurations consists of reporting criteria and format. In 5G, they also contain reference signal type that indicate the reference signal UE uses for beam and cell measurements. Reporting criteria is what triggers UEs to send a measurement report which can be periodic, or event based (for example neighbor cell's signal getting better than some threshold). Reporting format specifies quantities, e.g., numbers of cells, that UE includes in the measurement report. The reporting configuration could also indicate reporting of Cell Global Identifier (CGI) to get full identifier of a cell. Finally, a measurement object is linked to a reporting configuration by what is called a measurement identifier that identifies a measurement configuration. Multiple measurement identifiers can be used to link several measurement objects to several reporting configurations.

Next, the UEs measure and report according to the measurement configuration in an RRC message called measurement report. The standards mandate that this message is sent by UE only after successful security activation. It means that the message is encrypted, and integrity protected so that unauthorized parties or sniffers cannot read or modify the reports sent by the UEs. The report consists of measurement results identified with measurement identifier so that they can be linked to corresponding measurement configuration. As per measurement configuration, the report may further consist of measurements on serving and neighbor cells such as physical cell identifiers (PCI), received signal received power (RSRP), received signal received quality (RSRQ), signal to interference plus noise ratio (SINR).

As shown in Fig. \ref{fig:rrc_proc}, there is yet another procedure which can also be used by the RAN to configure the capable UEs to collect logs even during idle or inactive states. It is called logged measurement configuration. For the UEs which were configured to perform logged measurements, the RAN can obtain the reports when the UEs come back to connected state by sending RRC message called information request, and UEs will return reports in an information response message. We will not go into further specifics for brevity. Detail information are available in 3GPP TS 36.331 \cite{3gpp36331} for 4G, 3GPP TS 38.331 \cite{3gpp38331} for 5G, see clause 5.5 on Measurements.

\subsubsection{Data Fetch Procedures}
The Analyzer engages using these procedures to obtain data from Data Collectors and the Auxiliary data. These procedures are rather logical and can be materialized in numerous, non-exclusive, ways as follows. One way is to use application layer protocols over TCP/IP connections such as SSH File Transfer Protocol (SFTP). Another way is to use a Network Attached Storage (NAS) or a database server. Tools like Rsync and Secure Copy (SCP) can also be used.

As has been mentioned earlier, Analyzer can be deployed as a part of RAN. In such cases when base stations play the role of both the Data collectors and Analyzer components of Murat, this data fetch procedures become internal to base stations.

\subsection{Analysis Step}
\begin{figure}[h]
\centering
\includegraphics[width=2.5in]{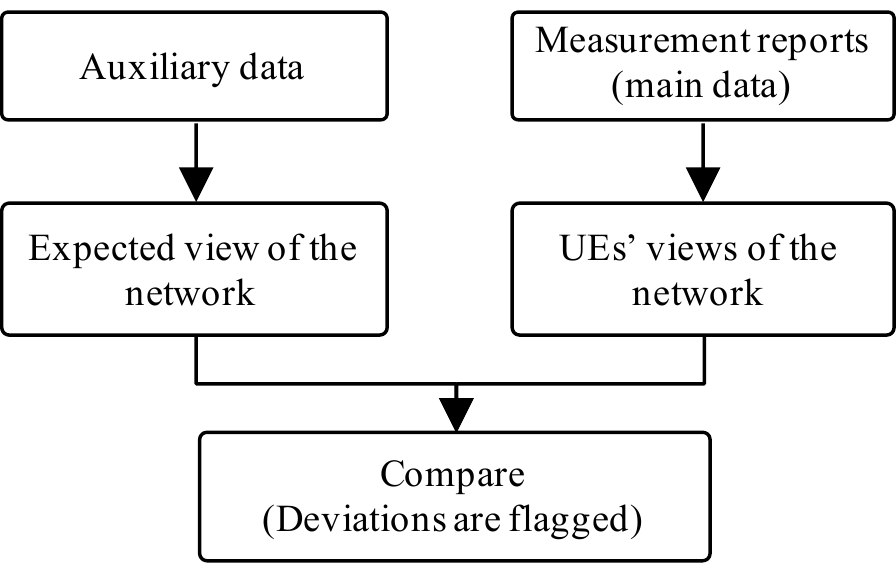}
\caption {Overview of Analysis step} 
\label{fig:analysis_steps} 
\end{figure}

The Analysis step is performed by the Analyzer component of Murat. The overall goal of this step is to identify the presence of false base stations in a mobile operators' network. A high-level overview of the Analysis step is illustrated in Fig. \ref{fig:analysis_steps}. Measurement reports are the main data in the Analysis step. They are used to identify the views of the network from the UEs' perspective, e.g., how many and what types of cell they observed in a certain area. Auxiliary data, on the other hand, are used to form the expected view of the network, e.g., how many and what types of cells are in fact expected to be present in that area. These two views are compared and if the UE's views deviate from the expected view, then such deviations are flagged. 

One or more functions of the Analyzer component  are executed in the Analysis step. Data processor function in Analyzer prepares the data for analysis by parsing the measurement reports obtained from Data collectors and Auxiliary data. 

Data prepared by the data processor is then run through one or more strategies to detect false base stations. One strategy is to use a rule-based algorithm that examines selected fields in the measurement reports and flags suspicious neighbor cells as false. Some examples follow. We want to note that following rules are only examples to give an idea of what type of rules may be suitable. They do not reflect universal rules suitable for all networks.  

Example rules for intuition:
\begin{itemize}
    \item	\textbf{Setting}: Only the PCIs between 0 and 450 are allocated to legitimate cells in a region; \textbf{Rule}:\textit{ PCIs in range 0-450 $\rightarrow$ legitimate cell; otherwise $\rightarrow$ false cell }
    
    \item   \textbf{Setting}: PCIs between 400 and 410 are not allocated to any legitimate cells in a region; \textbf{Rule}:\textit{ PCIs in range 400-410 $\rightarrow$ false cell; otherwise $\rightarrow$ legitimate cell }
  
    \item   \textbf{Setting}: PCIs 312, 313, and 314 are lone cells in a remote region with no neighboring cells; \textbf{Rule}:\textit{ PCIs reported together with 312, 313, and 314  $\rightarrow$ false cell }

    \item   \textbf{Setting}: All cells other than 263 in a region are put to sleep during non-office hours; \textbf{Rule}:\textit{ PCIs other than 263 reported between 18:00-8:00 $\rightarrow$ false cell }

    \item   \textbf{Setting}: Historical data in a region show that signal strengths received by UEs is always less than -60 dBm; \textbf{Rule}:\textit{ RSRP $<$ -60 dBm $\rightarrow$ legitimate cell; otherwise $\rightarrow$ false cell }

    \item   \textbf{Setting}: Historical data in a region show that signal qualities received by UEs is never greater than -9 dB; \textbf{Rule}:\textit{ RSRQ $>$ -9 dB  $\rightarrow$ false cell; otherwise $\rightarrow$ legitimate cell }

    \item   \textbf{Setting}: All 2G cells are phased out from a region; \textbf{Rule}:\textit{ 2G/GERA cells $\rightarrow$ false cell; otherwise $\rightarrow$ otherwise check other rules  }

    \item   \textbf{Setting}: All 3G cells are phased out from a region; \textbf{Rule}:\textit{ 3G/UTRA cells $\rightarrow$ false cell; otherwise $\rightarrow$ otherwise check other rules  }

    \item   \textbf{Setting}: Only three mobile network operators with codes 11, 12, and 13 are allowed to operate in a country; \textbf{Rule}:\textit{ mobile network code among (11,12,13)  $\rightarrow$ legitimate cell; otherwise $\rightarrow$ otherwise check other rules  }

\end{itemize}
The ranges, thresholds, and other parameters mentioned above can be either hard-coded in rules or taken as input from customization parameter as part of Auxiliary data. 

Rule-based strategy works particularly well when the relevant parameters are known clearly and accurately, for example, if a 2G network has been phased out, there cannot be any 2G cells visible to UEs. In those cases, this strategy is simple to implement. So, rule-based strategy can be the first in a detection chain before executing other strategies to detect false base stations. 

There could be other cases when rules are impractical because there are no reasonable parameters. For example, when legitimate cells in a wider region are considered, the RSRP/RSRQ values could vary from min to max of the allowed range and all the allowed PCI values may be used. In those cases, more intelligent strategies should be put into place like the ones using machine learning (ML) algorithms. Such strategies are left for future research. 

In the following sections, we describe how we realized Murat and validated it by doing lab experiments and operator trial.

%% file: parts/4.lab_experiments.tex
\section{Lab Experiments}
\label{labex}
\subsection{Environment}
\label{labex_env}
We tested Murat in one of our test labs. The main purpose was to verify that it is possible to detect false base stations of different generations even though the network part of Murat is only implemented using measurement reports received by a 4G base station.  

Our lab experiment setup consisted of an ordinary UE and a legitimate 4G network with one 4G base station (eNB) operating multiple 4G cells. The UE connected to this legitimate base station and browsed the Internet regularly.

There were two other standalone base stations (one cell each), one of which was a 4G eNB and another was a 2G BTS. They were standalone in the sense that they were not connected to any core network and were merely broadcasting cell information to announce their presence. The standalone 4G eNB was operating a 4G cell with the PCI 204 which was not used in the legitimate 4G network.

Expected outcome of this experiment was that Murat detects the standalone 4G cell as false because a 4G cell with that PCI 204 was not supposed to be present. It was also expected that Murat detects the standalone 2G cell as false because the network was not operating any 2G cells.

\begin{figure}[h]
\centering
\includegraphics[width=3.5in]{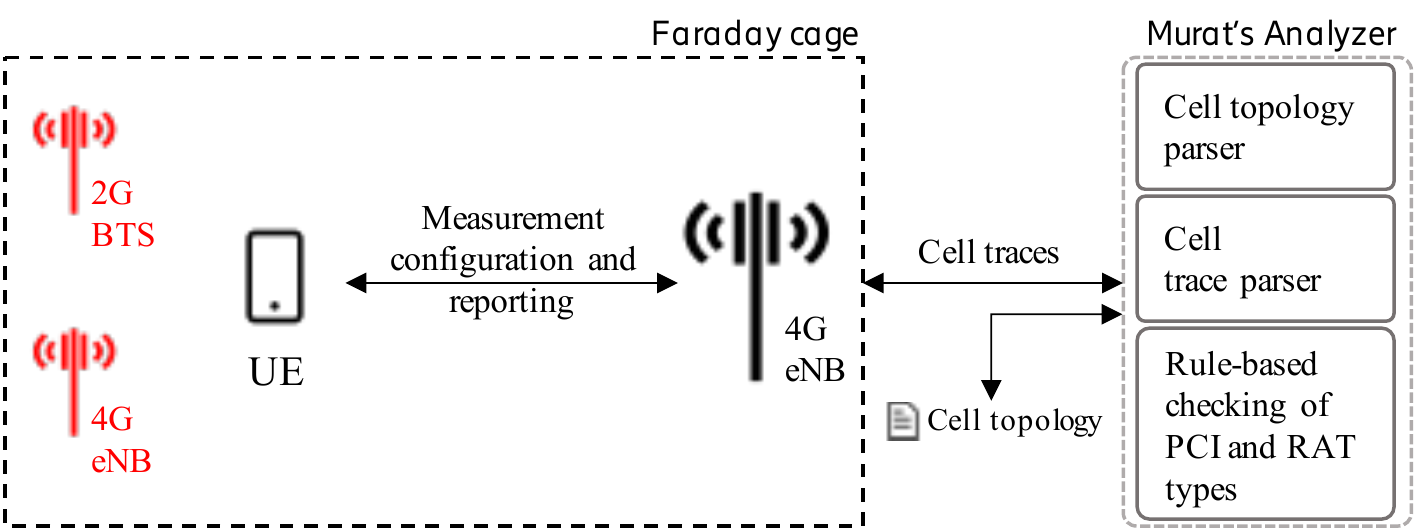}
\caption {Lab experiment setup} 
\label{fig:lab_exp_setup} 
\end{figure}

The lab setup is shown in Fig. \ref{fig:lab_exp_setup}. We used a Faraday's cage such that the UE was physically put in the Faraday's cage while radio from base stations (technically cells) were fed in via RF cables. 

In this setup, the UE was the Murat's Probe, the legitimate 4G eNB was its Data collector, and a separate server was its Analyzer. In the following sections, we describe the data collection and analysis steps in Murat.

\subsection{Data Collection Step}
\label{labex_dc}
The data collection step comprised of data reporting procedures between UE and the 4G eNB, and data fetch procedures between the 4G eNB and the Analyzer. 

Standard 3GPP RRC procedures as described earlier in Section \ref{section:meas_report} were used as the data reporting procedures. Both 4G and 2G measurement configurations were sent by the 4G eNB to the UE. 
Complete measurement configurations (with various information elements) from our experiments are shown in Fig. \ref{fig:4g_meas_conf} and Fig. \ref{fig:2g_meas_conf} for sake of completeness. We do not explain all the information elements for brevity; readers are referred to \cite{3gpp36331} for details. Information elements relevant for this paper are discussed below.

Fig. \ref{fig:4g_meas_conf} shows an example 4G configuration. This configuration is identified by measurement identifier 4 which further links together the measurement object 1 and reporting configuration 4. Among others, carrier frequency value 100  indicates E-UTRA Absolute Radio Frequency Channel Number (EARFCN) 100, which means band 1; and event A3 means an event when a neighbor cell becomes amount of an offset better than the serving cell.

\begin{figure}[h]
\begin{center}
\begin{lstlisting}
RRC {
  pdu value DL-DCCH-Message ::= {
    message c1 : rrcConnectionReconfiguration : {
     rrc-TransactionIdentifier 1,
     criticalExtensions c1:rrcConnectionReconfiguration_r8:{
        measConfig {
          measObjectToAddModList {
            MeasObjectToAddMod {
              measObjectId 1,
              measObject measObjectEUTRA : {
                carrierFreq 100,
                allowedMeasBandwidth mbw6,
                presenceAntennaPort1 FALSE,
                neighCellConfig '10'B,
                offsetFreq dB0}}},
          reportConfigToAddModList {            
            ReportConfigToAddMod {
              reportConfigId 4,
              reportConfig reportConfigEUTRA : {
                triggerType event : {
                  eventId eventA3 : {
                    a3-Offset-6,
                    reportOnLeave FALSE},
                  hysteresis 2,
                  timeToTrigger ms40},
                triggerQuantity rsrp,
                reportQuantity both,
                maxReportCells 4,
                reportInterval ms480,
                reportAmount r1}}},
          measIdToAddModList {
            MeasIdToAddMod {
              measId 4,
              measObjectId 1,
              reportConfigId 4
}}}}}}}
\end{lstlisting}
\end{center}
\caption {Example 4G measurement configuration} 
\label{fig:4g_meas_conf} 
\end{figure} 

Similarly, Fig. \ref{fig:2g_meas_conf} shows an example 2G configuration in which we additionally illustrate how the 4G eNB, after knowing that some 2G cell exists, can instruct the UE to measure addition information on that 2G cell. In this example, the 2G configuration identified as measurement identifier 3 is requesting a Cell Global Identifier (CGI) for a particular 2G cell. The CGI is the globally unique identifier of a cell which also consists of Mobile Country Code (MCC) and Mobile Network Code. 

The eNB was configured to save the measurement reports received from the UE as log files, also called cell traces. As a part of data fetch procedure, these cell traces were fetched from the eNB using our internal tool and saved in a shared network folder accessible to the Analyzer. Additionally, we created a cell topology file in a CSV format and saved in the shared network folder. This cell topology file, which was the Auxiliary data, contained MCC, MNC, RAT types, and cell identifiers for the legitimate cells in the network. In this setup, there were only 4G RATs that were legitimate.

\begin{figure}[h]
\begin{center}
\begin{lstlisting}
RRC {
  pdu value DL-DCCH-Message ::= {
    message c1 : rrcConnectionReconfiguration : {
    rrc-TransactionIdentifier 3,
    criticalExtensions c1:rrcConnectionReconfiguration_r8:{
        measConfig {
          measObjectToAddModList {
            MeasObjectToAddMod {
              measObjectId 2,
              measObject measObjectGERAN : {
                carrierFreqs {
                  startingARFCN 12,
                  bandIndicator dcs1800,
                  followingARFCNs explicitListOfARFCNs:{}},
                offsetFreq 0,
                ncc-Permitted '11111111'B,
                cellForWhichToReportCGI {
                  networkColourCode '111'B,
                  baseStationColourCode '011'B}}}},
          reportConfigToAddModList {
            ReportConfigToAddMod {
              reportConfigId 3,
              reportConfig reportConfigInterRAT : {
                triggerType periodical : {
                  purpose reportCGI},
                maxReportCells 1,
                reportInterval ms1024,
                reportAmount r1}}},
          measIdToAddModList {
            MeasIdToAddMod {
              measId 3,
              measObjectId 2,
              reportConfigId 3
}}}}}}}
\end{lstlisting}
\end{center}
\caption {Example 2G measurement configuration (with CGI request for a particular 2G cell)} 
\label{fig:2g_meas_conf} 
\end{figure}

\subsection{Analysis Step}
\label{labex_an}
This step was performed at the Analyzer in a separate server as follows. One of the functions in the Analyzer parsed the cell topology which was the Auxiliary data stored in a shared network folder as a CSV file. Another function parsed the cell traces which were also stored in that folder using an internal tool. From the parsed cell traces, RRC measurement reports for the 4G RAT (E-UTRA) and the 2G RAT (GERA) were obtained. 

Similar to measurement configurations, complete measurement reports from our experiments are shown in Fig. 9 and Fig. 10 for sake of completeness. However, we discuss only the information elements relevant for this paper.

An example of 4G report is shown in Fig. \ref{fig:4g_meas_rep}. The \textit{measResultListEUTRA} is part of the measurement identifier 4 which relates to the 4G configuration discussed earlier. The UE reported two neighbor 4G cells with their PCIs and signal properties. Next, a rule-based strategy was used to check if the reported PCI and RAT types were as expected or not. The parsed measurement reports were compared against the parsed cell topology. The topology did not have any 4G cell that was assigned the PCI 204, therefore, the reported 4G PCI 204 was flagged as belonging to a false base station. 

\begin{figure}[h]
\begin{center}
\begin{lstlisting}
RRC {
  pdu value UL-DCCH-Message ::= {
    message c1 : measurementReport : {
      criticalExtensions c1 : measurementReport_r8 : {
        measResults {
          measId 4,
          measResultPCell {
            rsrpResult 41,
            rsrqResult 33},
          measResultNeighCells measResultListEUTRA : {
            MeasResultEUTRA {
              physCellId 204,
              measResult {
                rsrpResult 41,
                rsrqResult 2}},
            MeasResultEUTRA {
              physCellId 366,
              measResult {
                rsrpResult 41,
                rsrqResult 5
}}}}}}}}
\end{lstlisting}
\end{center}
\caption {Example 4G measurement report} 
\label{fig:4g_meas_rep} 
\end{figure}

Similarly, 2G measurement reports were received and since the topology did not have any 2G cell at all, the 2G cell in the measurement reports was also flagged by the rule-based strategy. Fig. \ref{fig:2g_meas_rep} shows the UE's 2G report in response to the measurement configuration shown earlier in Fig. \ref{fig:2g_meas_conf}. The \textit{measResultListGERAN} is a part of measurement identifier 3 and, as was asked, the UE also reported the neighbor cell's full CGI so that it is known what country and network codes the false base station was using for the 2G cell.

\begin{figure}[h]
\begin{center}
\begin{lstlisting}
RRC {
  pdu value UL-DCCH-Message ::= {
    message c1 : measurementReport : {
      criticalExtensions c1 : measurementReport_r8 : {
        measResults {
          measId 3,
          measResultPCell {
            rsrpResult 44,
            rsrqResult 31},
          measResultNeighCells measResultListGERAN : {
            MeasResultGERAN {
              carrierFreq {
                arfcn 12,
                bandIndicator dcs1800},
              physCellId {
                networkColourCode '111'B,
                baseStationColourCode '011'B},
              cgi-Info {
                cellGlobalId {
                  plmn-Identity {
                    mcc {
                      MCC-MNC-Digit 1,
                      MCC-MNC-Digit 1,
                      MCC-MNC-Digit 1},
                    mnc {
                      MCC-MNC-Digit 1,
                      MCC-MNC-Digit 1}},
                  locationAreaCode '0000000000000001'B,
                  cellIdentity '0000000001101111'B}},
              measResult {
                rssi 63
}}}}}}}}
\end{lstlisting}
\end{center}
\caption {Example 2G measurement report} 
\label{fig:2g_meas_rep} 
\end{figure}

%% file: parts/5.operator_trial.tex
\section{Operator Trial}
\label{optr}
\subsection{Environment}
\label{optr_env}
\begin{figure}[h]
\centering
\includegraphics[width=3.5in]{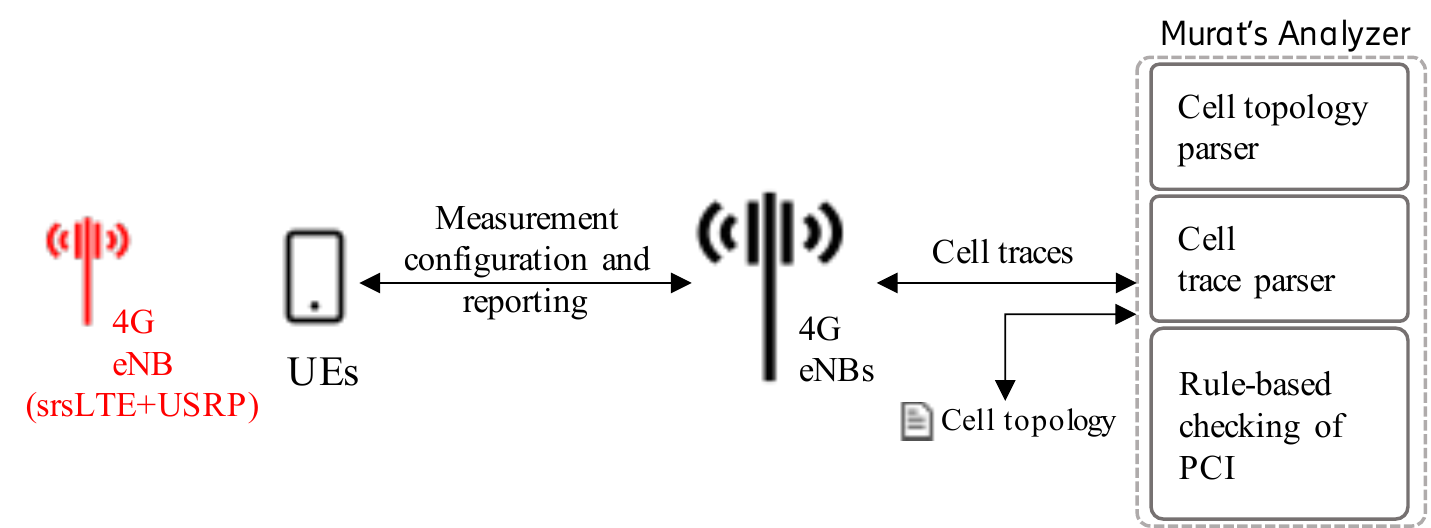}
\caption {Operator trial environment} 
\label{fig:op_trial} 
\end{figure}

After the lab experiment, we collaborated with a real operator and conducted trial in their 4G network. The trial environment is shown in Fig. \ref{fig:op_trial}. The basic components in the operator trial are same as the lab experiment, since both adhere to Murat's design. So, we will only point out important peculiarities in the operator trial.

The trial was conducted in an open area without using any Faraday’s cage and time chosen by the operator where multiple legitimate base stations operated. Users would pass by that area, meaning that the UEs (acting as Probes) were ordinary UEs. The trial had only one standalone base station acting as a "planted" false base station and it was used solely for the purpose of test.  The "planted" 4G eNB was assembled using publicly available tools. Its hardware part consisted of USRP B210 \cite{usrp_b210}, a software defined radio (SDR) from Ettus Research \cite{ettus}, and its software part was the open-source LTE software suite called, srsLTE \cite{srsLTE}. These components are popular among researchers working with false base station attacks. The planted 4G eNB was turned on couple of times with specific PCI values that were not in use by the operators' legitimate base stations in the closest vicinity. We configured the parameters called \textit{dl\_earfcn}, \textit{mcc}, \textit{mnc}, \textit{tac}, \textit{enb\_id}, \textit{cell\_id}, \textit{phy\_cell\_id}, and executed eNB part of the srsLTE as-is. We did not make any modifications in these hardware or software components. Special cares were taken when operating the planted 4G eNB. Whenever it was turned on, it was turned on only for some minutes. Further, no attack was performed by the planted 4G eNB other than announcing itself as a cell belonging to the operator's network. It was very important to do so in order to minimize any unintentional inconvenience to users nearby.

Expected outcome of this trial was that Murat detects all the PCI values announced by the planted 4G eNB as false because none of them were supposed to be present in the location where it was operated.  

\subsection{Data Collection Step}
\label{optr_dc}
The data reporting and fetch procedures were similar to the lab experiment except that the measurement configuration in the operator's network was untouched. The cell trace files contained measurement reports from legitimate eNBs in the area where our planted 4G eNB was placed and during the time covering its operation. A summary of data collection is given in Table ~\ref{table:datacollection}.


\begin{table}[h]
\caption{Summary of data collection}
\label{table:datacollection}
\centering
\begin{tabular}{|c||c|}
\hline
number of eNBs  & 5\\
\hline
total number of 4G cells & 36\\
\hline
total number of 4G measurement reports & 7739\\
\hline
number of unique PCIs collected & 156\\
\hline
\end{tabular}
\end{table}

\subsection{Analysis Step}
\label{optr_an}
Similar to the lab experiment, the Analysis step was performed in a separate server and a rule-based strategy was used. However, we only considered the 4G RAT and therefore it was only the reported PCIs that were checked. 

\begin{figure}[h]
\centering
\includegraphics[width=3.5in]{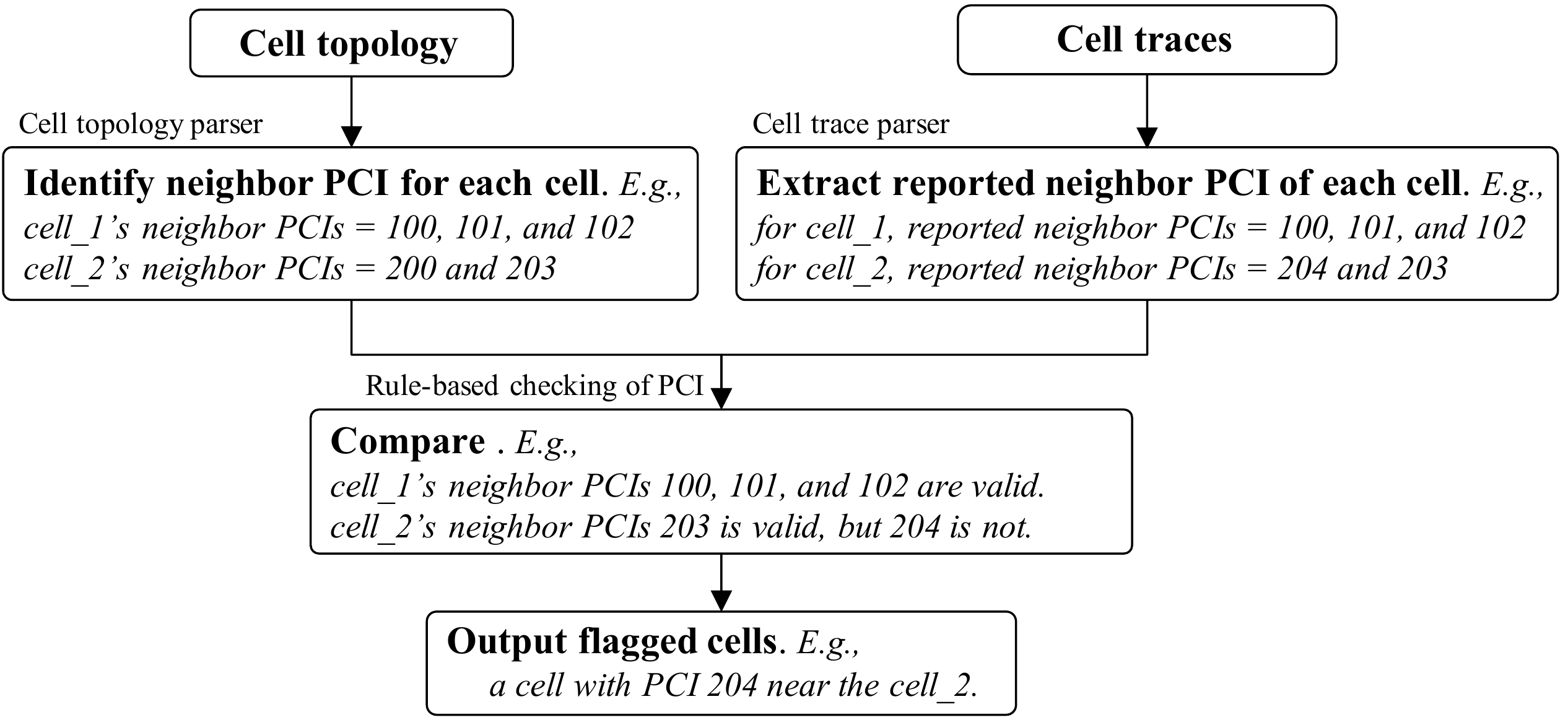}
\caption {Illustration of Analysis step in the operator trial} 
\label{fig:analysis_trial} 
\end{figure}

Some extra treatments were also needed in the real operator network which we describe next. Fig. \ref{fig:analysis_trial} depicts the overall analysis step in the operator trial. 

Unlike in the lab experiment, a strategy to flag a cell with a certain PCI would not work when that PCI does not exist in the cell topology. It is so because the total number of PCIs available for use is only few (504 in 4G) and they are only locally unique. Since a real operator network would have more cells than those, all the PCIs would be used at least in some region and would appear in the cell topology. Hence, in the real operator network, the cell topology parser first needed to identify neighbor PCIs for each cell.

The topology file that we received from the operator did not contain actual geolocation of cells, meaning that there was no obvious metric to use for defining neighbor cells. However, it contained handover relations between cells. When we did an initial test of Murat using similar topology file and considering only the cells with handover relations as legitimate neighbors, Murat had flagged 63 neighbor PCIs as false neighbors. These were false positives -- since we expected 0 false neighbors -- which can be explained by the fact that although the handover relations could be used to identify some neighbor cells, they are not sufficient to identify all neighbor cells.  The reason is -- even if a neighbor cell is geographically close to a cell, that neighbor cell does not necessarily also have a direct handover relation with the cell. In other words, it is practically possible that a direct handover relation does not exist between two geographically nearby cells because there is some another cell which is closer to both of them and is the handover candidate for both. 

In order to address the above issue identified during initial testing, we devised a technique to identify which cells are geographically close to each other. In this technique, for each cell, we defined neighbors to be combination of cells with immediate handover relations plus cells with handover relation after a N hops. In other words, N-neighbor-of-neighbor was also considered as a neighbor. Using this approach in our initial test resulted in a sharp decrease in false positives to 1 with N = 1, and to 0 with N = 2. Then, we tested this approach using the topology file for the trial and got 0 false positives with N = 1. We note that the choice of N is not strict; its effectiveness and limitations are discussed in Section \ref{eff}.E.

Rest of the Analysis step was to compare the parsed measurement reports against the parsed cell topology containing each cell's identified neighbor PCIs. All the PCIs that the planted 4G eNB was operating on were flagged. 

%% file: parts/6.standardization.tex
\section{Standardization}
\label{std}
We first gave our input \cite{S3-170463} to 3GPP's group responsible for security during the study phase of 5G security. Later, when the first set of 5G specifications (called Release 15) were getting finalized, we proposed to include our approach in the formal 5G security standard. After also getting vetted \cite{S3-171568, R1-1711997, R2-1709980, R4-1711318}  by other groups in 3GPP who work with RAN aspects, it was adopted into an informative annex of the 5G security specification 3GPP TS 33.501 \cite{3gpp33501}.

%% file: parts/7.effectiveness_limitations.tex
\section{Effectiveness and Limitations}
\label{eff}

In this section, we revisit some aspects and discuss Murat's effectiveness as well as limitations that come either from Murat's design or specific implementation choices. 
\subsection{Data collection}
\label{eff_dc}

Since Murat fundamentally relies on measurement reports from UEs, it functions as long as at least one or few UEs are connected to the legitimate network. If a false base station lures all UEs so that no UE sends measurement reports to the legitimate network, then Murat will not function.  We note that a false base station acting in this manner for a long period of time would result in that no UEs connect to the legitimate base stations in the area. In a relatively well populated area this would in itself be an indication of that something in the area is misbehaving and that an investigation of the cause is necessary. 

Measurement reports used in Murat are part of standard 3GPP RRC messages which are encoded in ASN.1 Unaligned Packed Encoding Rules (UPER) format. Therefore, parsing those raw measurement reports would work gracefully even in a multi-vendor deployment of a mobile operator’s network. Nevertheless, some integration efforts, may be necessary to handle cases if some vendors’ base stations provide those reports with proprietary flavors, e.g., already decoded measurement reports in formats like JSON or CSV, with or without compression.

Cell topology data with information on legitimate base stations is an important Auxiliary data that is useful in the analysis. Even though there are some 3GPP standards [1] on data format like XML, it is only likely that different vendors have their own extensions, therefore, cell topology parser would need to adapt accordingly. It is also likely that parsing other Auxiliary data like customization parameters and network event logs would need vendor specific adaptations.

\subsection{Multi-RAT detection}
\label{eff_Multirat}
Our lab experiment and operator trial prove that the general principle of using unmodified ordinary UEs to report their views and comparing it to the network's view works. An important aspect that was also shown in the lab experiment is even though a UE is connected to a specific 3GPP RAT, it can still be asked to report measurements for other 3GPP RATs. Hence, in principle, mobile network operators could deploy the Data collector component of Murat in one of any generations and detect false base stations on all generations. Note that the Analyzer component of Murat can be same regardless of which generation the Data collectors are deployed in.

Nevertheless, it is important to note that the 3GPP standards currently do not allow a 5G network to directly interwork with a 2G network for reasons of security isolation among others. It means that a UE in a 2G network cannot directly switch to a 5G network and vice-versa. Consequently, 2G measurements are not specified in 5G network's measurement report and vice-versa. The same is true for a 3G network but in one direction, i.e., while a UE can switch from a 5G network to a 3G network, the opposite is not allowed. At the time of writing, it is only the 4G network that can fully interwork with all generations. Therefore, as of today, a mobile network operator would gain optimal benefit from Murat by deploying its Data collector to collect measurement reports from 4G since measurement reports in a 4G RAT can contain all 2G, 3G, 4G, and 5G measurements. However, there could be cases when the Data collector is deployed only for a single generation/RAT that Murat does not detect false stations in another generation/RAT. In case of the Data collector is deployed only for a 4G RAT and in a certain area/time, all UEs support only 2G RAT, e.g., configured to operate on 2G-only mode. Since there are no UEs connected to the operator's 4G network in that area/time, Murat will not receive any measurement reports at all in that area/time and a 2G false base station will go undetected even though operational. We consider this as a very rare and improbable case since it is not likely that all users configure their UEs to use a lower generation (and slower) network. It could also be other way round that all the UEs support (or have enabled) only the 4G RAT in a certain area/time. In this case, Murat will not get any 2G RAT measurement reports, meaning that 2G false base stations in that area/time will not be detected. We consider this to be arguably fine because with no UEs supporting 2G RAT, there is no victim and therefore the 2G false base station should be almost useless for the attacker.

\subsection{Operator specific detection}
\label{eff_op}
 When deployed in one operator's network, Murat detects false base stations running on frequencies operated by that operator but not in those operated by other operators.

\subsection{Rule-based strategy}
\label{eff_rule}

Rule-based strategy of checking PCIs, which we implemented, works well in scenarios when the attacker is forced to operate its false base station with a PCI not used by the legitimate base stations (e.g., to maintain a stable attack as indicated in \cite{rupprecht-19-layer-two}). However, it could be possible in principle that a resourceful attacker uses some advanced techniques to evades this rule-based detection strategy by operating on same PCIs used by the legitimate cells and still maintain stable attack. In order to detect such false base stations, different strategy utilizing other innate properties of radio communication, e.g., signal strength and quality, would be required and is a topic for a separate study.

We want to stress that static rules are not suitable when the cell topology changes frequently, due to operators using features like automatic neighbor relations (ANR). Manual maintenance of these rules is not practical neither, especially for large scale deployments. Therefore, when a rule-based strategy is used, the rules should be implemented such that they automatically reflect up-to-date cell topology information, e.g., the rules being updated from a (near) real-time database.

\subsection{Neighbor-of-Neighbor technique}
\label{eff_neig}
The neighbor-of-neighbor technique that we implemented in the operator trial served its purpose (see Section \ref{optr}.C). However following points should be kept in mind. 

Neighbor-of-neighbor cell information must take frequencies into account to be effective. This is because even though a PCI uniquely identifies a cell in a given area operating on a specific frequency,  the same PCI may legitimately be used to identify a different cell in the same area, but one that operates on a different frequency. Consequently, considering any cell with a certain PCI as an acceptable neighbor on all frequencies, can lead toa false negative when a false base station uses the PCI of a legitimate neighbor but on a different frequency. For example, a PCI X could be a neighbor in frequency f1 but not in frequency f2. Considering the PCI X to be a neighbor in f2 would lead to false negative if a false base station uses the PCI X in f2.

Increasing the hop count to high values (e.g., neighbor-of-neighbor-of-neighbor or more) means that even the cells which are geographically far could end up being considered as valid neighbors. This defeats the concept of neighbor and would produce false negatives. 

Further, although unlikely, it cannot be ruled out that there won’t be any standalone legitimate cell in a neighborhood which has no handover relation with any other cell whatsoever. Such standalone cells would produce false positives. 

Therefore, using the actual geolocation of cells as a first choice to identify neighbors would be a better way. 

%% file: parts/8.future_work.tex
\section{Future Research} \label{fr}
Our multi-RAT false base station detector could be augmented with additional features like below. Each of them deserves research topics on its own, nevertheless, we give some initial pointers for interested researchers.

\subsection{Smart distribution of measurements}

A rewarding feature to add would be a smart distribution of measurement load across UEs. Although the network can practically just keep its existing measurement configuration setup and analyze existing data, it could benefit more by being able to actively setup new configurations, e.g., by choosing timings and location for collection of particular type of measurements to best fit its detection needs. However, it is important to keep the demand or effect on UEs to minimum in terms of service interruption or battery consumption. This is where the smart distribution could be beneficial, for example, by being able to distribute the measurement reporting load across the UEs by partitioning the cells to be measured and then configuring each UE to provide measurement reports for the cells only in one of these partitions. 

\subsection{Enriched measurement reports}

Another beneficial feature would be enriched measurement configuration and reporting. By enriched, we mean support for collection of new information about camped and neighbor cells that could enhance detection of false base stations attacks. For example, state of the system information broadcast messages received by UEs could be analyzed in the network to detect if an attacker tampered some information like cell barring, support for IP Multimedia Subsystem (IMS) emergency, system information scheduling, and neighbor cell list. Additionally, logged measurements could be enriched so that collection of following information is supported: number of reject messages the UE had received, presence of erratic radio signal not associated with any normal reference signals, presence of signal associated with normal reference signals but without any readable system information, presence of signals associated with normal reference signals and with system information but with the wrong information making it impossible to access the network. 

\subsection{Machine learning based detection}
Machine learning based approaches on measurement reports should be investigated. Since false base stations usually transmit at higher-than-normal signal strength in order to lure the nearby UEs, they would most likely induce inherent change in the surrounding radio environment. Hence, it is worth doing research on if machine learning models could be trained to detect disturbances induced by the false base stations in terms of signal properties.

\subsection{Post-detection actions}
There could also be research on reactive actions (protocol level or otherwise) to be taken by operator after detection of false base station. This feature may be a mix of implementation specific and standardized feature, e.g., automatic positioning and containment of detected false base stations.

%% file: parts/9.conclusion.tex
\section{Conclusion}
\label{conc}
We presented Murat, a network-based false base station detector, which is capable of detecting false base stations operating in multiple 3GPP Radio Access Technologies (RAT). By validating it in a lab  experiment and a real operator trial, we showed that Murat does not require any modification to mobile phones and can work even if data is collected from only one type of RAT in mobile operators' network. These make Murat more effective than other detection systems that either rely on special software on the mobile phones or cover false base stations operating in a single RAT. We also discussed practical insights for a real-world deployment of Murat. Murat's approach was proposed to 3GPP and was adopted in the mobile network standards.

%% file: parts/91.acknowledgement.tex
\section*{Acknowledgment}
This paper is a result of many years of work and contributions on various fronts from our several colleagues . We thank Per Bardun, Daniel Glifberg, Antti Jaakkola, Anna Kåhre, Irfan Bekleyen, Yasin Yur, Beste Akkuzu, Oscar Ohlsson, Angelo Centonza, and Marc Mowler.

This work was partially funded by The Scientific and Technological Research Council of Turkey (TUBITAK), under 1515 Frontier RD Laboratories Support Program with project no:5169902.

This work was partially supported by the Wallenberg AI, Autonomous Systems and Software Program (WASP) funded by the Knut and Alice Wallenberg Foundation.